\documentclass[aps,amsmath,amssymb,pre,11pt,onecolumn]{revtex4}
\usepackage{graphicx}
\usepackage{natbib}
\usepackage{multirow}
\usepackage{sidecap}
\usepackage{times}

\usepackage{color}

\begin{document}

\title{Self-healing Dynamics of Surfactant Coatings on Thin Viscous Films}

\author{Stephen L. Strickland$^1$, Matthew Hin$^2$, M.~Richard Sayanagi$^3$, \\
Cameron Gaebler$^3$, Karen E. Daniels$^1$, Rachel Levy$^2$}

\affiliation{$^1$Dept. of Physics, NC State University, Raleigh, NC, USA \\
$^2$Dept. of Mathematics, Harvey Mudd College, Claremont, CA, USA \\
$^3$Dept. of Physics, Harvey Mudd College, Claremont, CA, USA}

\begin{abstract}
We investigate the dynamics of an insoluble surfactant on the surface of
a thin viscous fluid spreading inward to fill a surfactant-free region.
During the initial stages of surfactant self-healing, Marangoni forces drive an
axisymmetric ridge inward to coalesce into a growing central distension;
this is unlike outward spreading, in which the ridge only decays.
In later dynamics, the distension slowly decays and the surfactant
concentration equilibrates.
We present results from experiments in which we
simultaneously measure the surfactant concentration (using
fluorescently-tagged lipids) and the fluid height profile (via laser
profilometry). We compare the results to simulations of a mathematical model 
using parameters from our experiments.
For surfactant concentrations close to but below the critical monolayer
concentration, we observe agreement between the height profiles in the numerical
simulations and the experiment, but disagreement in the surfactant distribution.
In experiments at lower concentrations, the surfactant spreading and formation
of a Marangoni ridge are no longer present, and a persistent lipid-free region
remains. This observation, which is not captured by the simulations, has
undesirable implications for applications where uniform coverage is
advantageous. Finally, we probe the generality of the effect, and find that
distensions of similar size are produced independent of initial fluid
thickness, size of initial clean region, and surfactant type.
\end{abstract}

\maketitle


\section{Introduction \label{s:Intro}}

Scientists have been intrigued by the spreading of surfactants for centuries:
Benjamin Franklin famously wrote of sailors applying oil to the sea in order to
``calm the waters'' \citep{Franklin1774}. Quantitative experiments began with
the work of Agnes Pockels, whose letter to Lord Rayleigh was published in
\emph{Nature} \citep{Pockels1891} and describes the effect of kitchen powders on
surface tension. The techniques developed in their early work are alive today
\citep{Kaganer1999} in the form of Langmuir-Blodgett troughs used to study
molecular monolayers on fluid surfaces. Studies of surface tension
driven spreading began first on deep fluid layers \citep{Hoult1972}, but in more
recent decades, thin fluid films have been recognized as
underlying many complex biological and engineering processes. Applications include
pulmonary drug delivery \citep{Gaver1990}
and surfactant replacement therapy \citep{Gaver-1992-DST},
ocular surfactants and blinking dynamics \citep{Braun2012,Maki2010},
solute transport \citep{Engineering1994}, 
latex paint drying \citep{Evans2000, Gundabala2008, Gundabala2006},
ink-jet printing \citep{Hanyak2011}, and
secondary oil recovery \citep{Hanyak2012, Sinz2011c}.
In each of these processes, 
amphiphilic surfactant molecules relax the intermolecular bonds at the surface of an
underlying fluid and locally reduce the free energy. Gradients in the
concentration of insoluble surfactants cause gradients in the free energy,
known as Marangoni forces. These forces provide surface stresses that  induce
motion in the fluid and in turn advect the surfactant.  In the ideal case,
spreading results in a
homogeneous surfactant distribution, often the desired outcome for medical and
engineering applications.

The spreading dynamics of surfactants on thin liquid films are known to depend
on both the chemistry of the materials and the geometry of the system.
In many cases, the choice of an insoluble surfactant simplifies the
dynamics, since for low concentrations the transport of the surfactant molecules is therefore confined
to the surface of the fluid \citep{Craster2007}. A classic model by
\citet{Gaver1990} has long been used to predict the dynamics of
the fluid height profile and surfactant distribution for an insoluble surfactant
spreading on a thin fluid film. The model is based on the incompressible
Navier-Stokes equations, free surface and no slip boundary conditions, and
lubrication  theory. While theoretical
treatments of the subject \citep{Jensen1992a, Gaver-1992-DST,
Espinosa2013, shearer2006motion} have advanced our understanding, 
quantitative comparisons between theory and
experiments have only recently begun. In particular, predictions for the evolving
thickness of the fluid layer have been experimentally tested in both planar
and droplet geometries
\citep{Gaver-1992-DST,Dussaud1997,Dussaud2005,Fallest2010a,Hanyak2012}, but
much less is known about the quantitative dynamics of the surfactant
concentration \citep{Bull1999, Fallest2010a, Swanson2013}.

Two geometric properties, aspect ratio and interface curvature, are known to
play important roles in spreading dynamics. Surfactants spreading on fluid layers which are
thin enough to dewet are known to undergo fingering instabilities
\cite{Troian-1989-FIT,Hamraoui2004} that are not present when the fluid substrate is
thick enough to remain intact. Furthermore, the spreading of circular droplets
on thin films \citep{Dussaud2005,Fallest2010a,Bull1999,Swanson2013}
exhibit dynamics well-described by a radius which grows as $r(t) \propto
t^{1/4}$, while for thicker (non-lubrication) films the growth rate is $r(t) \propto
t^{3/4}$.  Curvature also affects the spreading rate; for thin fluid
films, a planar front spreads as $x(t) \propto t^{1/3}$ \citep{Hanyak2012},
while the aforementioned droplets spread outward at a rate of $r(t) \propto t^{1/4}$
\citep{Fallest2010a, Jensen1992a, Dussaud2005, Swanson2013}.

Astonishingly, a third possibility -- the spreading of a surfactant into a
depleted region -- has received comparably little attention.
Understanding more about
the dynamics in this geometry can shed light on the
interactions between disconnected or heterogeneous regions of surfactant
coverage as well as the
extent to which imperfectly-covered surfaces can self-heal.
Prior work by \citet{Jensen1994} analytically solved the model of
\citet{Gaver1990} using similarity solutions to understand the
dynamics leading up to the closure of the region at time $t_c$; the region is
predicted to shrink as $r(t) \propto (t_c - t)^{0.81}$. 
From a mathematical standpoint, this ``inward'' (or hole-closing) geometry
qualitatively
differs from ``outward'' and planar spreading in that the area to be covered
by surfactant is finite in first case and can be infinite in the later two.

\begin{figure}
\centering
\includegraphics[width=0.7\linewidth]{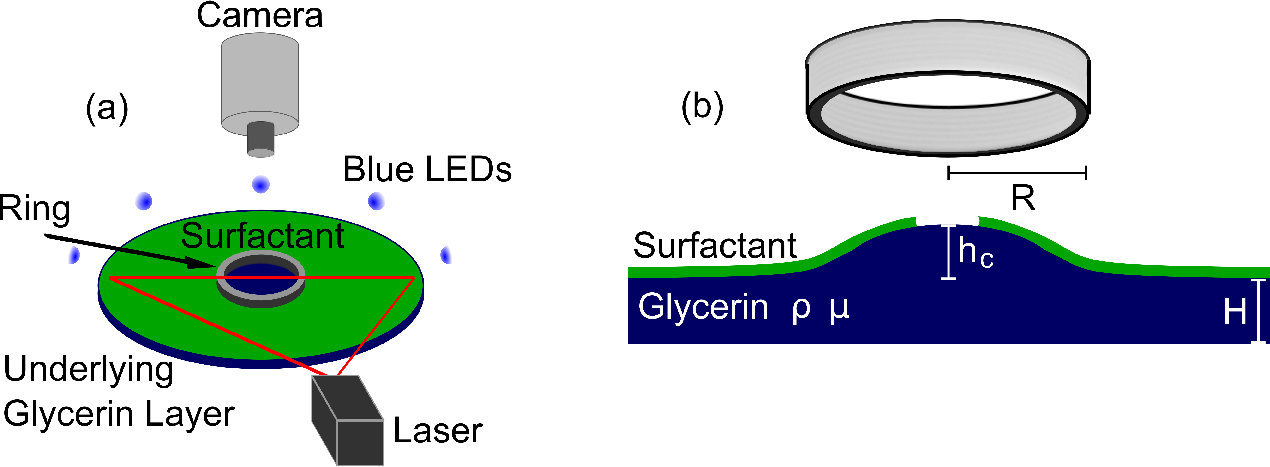}
\caption{[Color Online] {\bf Experiment schematic}  
($a$) experiment before the
barrier ring is lifted. ($b$) cross section of the fluid
and surfactant film after the barrier ring is lifted.}
\label{f:Schematic}
\end{figure}

In this paper, we address the inward spreading of surfactant into a previously
surfactant-free region using both experiment and modeling approaches, as
illustrated in Fig.~\ref{f:Schematic}. We
consider an initially flat fluid film partially covered with
surfactant; a circular region of radius $R$ is kept free of surfactant
while the area outside this region is coated with a uniform concentration of
surfactant $\Gamma_0$.  The initial fluid film thickness, $H$, is
millimetric, and is therefore small compared with the
several-centimeter masked region (lubrication theory applies).  Using this
geometry, we examine the
self-healing behavior of these regions, quantify the dynamics, and examine
to what extent the model by \citet{Gaver1990} is able to capture  these
behaviors.

For experiments and simulations performed under matched conditions, we report
the evolution of the height profile and surfactant concentration profile as the
surfactant spreads inward.
Interestingly, we observe that Marangoni forces 
initially drive fluid into the central region, creating a
distension which is also millimetric in scale.
By tracking the height of this distension, we are able to separate the dynamics
into a growth phase and a decay phase, and examine the dependence of the maximum
distension height on both geometric and material parameters of the system.
In addition, we simultaneously visualize the advancing front of surfactant by
using fluorescently-tagged lipids as the surfactant \citep{Fallest2010a}.
For large initial surfactant concentrations (approaching a monolayer in
coverage), the predicted and observed height profiles agree and the spreading
exponent associated with the leading edge of the surfactant matches
predictions from the self-similarity analysis of \citet{Jensen1994}. However,
the dynamics are approximately 10 times faster in the experiments than
in the simulations under the assumption that the glycerin is pure (anhydrous).
The leading edge is also much sharper in the experiments than in the
simulations. Finally, we observe that for initial surfactant concentrations
well below a monolayer in coverage, the hole does not completely heal on
experimental timescales. We close with a discussion of possible ways to
reconcile the observed differences. 

\section{Methods \label{s:methods}}

Using both experiments and simulations, we consider a thin fluid upon which
a surfactant spreads from an annular contaminated region into an initially
surfactant-free circular central region. This axisymmetric
geometry is shown schematically in Fig.~\ref{f:Schematic}. Because the
classic mathematical model \citep{Gaver1990} for  surfactant spreading on a
thin film makes predictions for
both the height profile $h(r, \theta, t)$ of the fluid film and the surfactant
concentration $\Gamma(r, \theta, t)$ on
its surface, we have designed and built an experimental apparatus which provides
access to both dynamics. Below, we describe the experimental methods and
materials, as well as the techniques for numerically solving the model in this
geometry.  

\subsection{Experiments \label{s:exper}}

The main apparatus consists of an aluminum well with radius $r_\mathrm{well}$ (either 11.1~cm or
14.6~cm) containing a millimetric layer of glycerin. In each experiment, we
place an initial concentration of surfactant $\Gamma_0$ in the region outside
of a cylindrical stainless-steel retaining ring, and this surfactant spreads inward after the
ring is lifted out of the fluid.  
The protocol is geometrically inverted from that used in prior experiments on outward spreading \cite{Bull1999, Gaver-1992-DST, Fallest2010a}.
We use two measurement techniques to capture the dynamics of this process:  laser profilometry
(LP), which measures the height profile of
the fluid surface, and fluorescence imaging (FI), which records the
spatial distribution of fluorescently-tagged surfactants. Using the LP
technique, we examine a broad range of surfactants (polydimethylsiloxane,
Triton X-305, sodium dodecyl sulfate, oleic acid, and NBD-PC
(1-palmitoyl-2-{12-[(7-nitro-2-1,3-benzoxadiazol-4-yl)amino]dodecanoyl} -sn-glycero-3-phosphocholine))
at bulk volumes in which the surfactant is no longer confined to a monolayer.
Using both FI and LP techniques, we examine a fluorescently-tagged
phosphocholine lipid (NBD-PC) at monolayer concentrations. The material
properties of the five surfactants are provided in Table~\ref{t:Surfs}.

The glycerin (Sigma-Aldrich) is initially $\geq 99.5\%$ anhydrous; at $20$ $^\circ$C,
viscosity $\nu = 14.1$~Poise \cite{Segur1951}, 
density $\rho = 1.26$~g/cm$^3$ \cite{Bosart1928},
and surface tension $\sigma_0 = 63$~dynes/cm \cite{Gallant1967}.
Because glycerin is both hygroscopic and temperature sensitive, the ambient
temperature and humidity, which ranged from $(22.8 \pm 0.4)^\circ$C and $19\% - 50\%$
respectively, can affect these physical parameters. In addition, it is possible that chloroform 
used during surfactant 
deposition might dissolve in the glycerin and decrease its viscosity. Simply considering 
hygroscopic effects, 
the material parameters could range as far as 
viscosity $\nu = 0.45$~Poise \cite{Segur1951}, 
density $\rho = 1.20$~g/cm$^3$ \cite{Bosart1928},
and surface tension $\sigma_0 = 65$~dynes/cm \cite{Gallant1967}.
Of these, viscosity is the most significant effect, and its decrease would also
cause the timescale of the dynamics to decrease (faster dynamics). 
In \S\ref{s:model}, we will introduce a parameter $\alpha$ to empirically correct the
redimensionalization in order to compare the simulations to the experiments, since the viscosity is unknown.

\begin{table} 
\centering
\begin{tabular}{ | c | c c c c | }
\hline
\multirow{2}{*}{Surfactant} & Surface Tension &
Solubility & \multirow{2}{*}{Solution} &
\multirow{2}{*}{$\Gamma_c$ $\frac{\mu g}{cm^2}$} \\
& $\sigma_s$ (dyne/cm) &  in glycerin & & \\
\hline
\multirow{2}{*}{NBD-PC \citep{Bull1999,Fallest2010a}}&
\multirow{2}{*}{35.5} &
\multirow{2}{*}{insoluble} & 1 $\mu$g of NBD-PC per & \multirow{2}{*}{0.3} \\
& & & $\mu$L solution in chloroform & \\
PDMS & 20.5 & slightly soluble & pure & -- \\
Triton X-305 & 49 & soluble & 1\% in water & -- \\
SDS \citep{Mysels1986} & 46  & soluble & 6 mM in water &  -- \\
oleic acid \citep{Gaver-1992-DST} & 32.79 &  insoluble & 0.1\% in
hexane
& 0.20 \\
\hline
\end{tabular}
\caption{{\bf Physical and chemical properties of the surfactants and surfactant solutions.}
For NBD-PC and oleic acid, the value of $\sigma_s$ corresponds to the
surface tension at or above $\Gamma_c$; at $\Gamma < \Gamma_c$, the surface
tension is higher.  For all other materials, $\sigma_s$ is the surface tension 
of the surfactant (or surfactant solution). 
Data for PDMS and Triton X-305 were obtained from
the manufacturer (Shin-Etsu and Dow Chemical, respectively).}
\label{t:Surfs}
\end{table}

\begin{table} 
\centering
\begin{small}
\begin{tabular}{|c|c|c|c|c|c|c|}
\hline
          & Well, & $R$ (cm)      & $H$ (mm)      & Surfactant   & $\Gamma_0$ ($\Gamma_c$) & $V$ ($\mu$L) \\
Technique & Base  & $\pm 0.04$~cm & $\pm 0.05$~mm &              & $ \pm 5\%$              &              \\ \hline \hline
FI \& LP1 & Si & 3.0 & 0.7 & NBD-PC & 0.2, 0.4, 0.6, 0.8, 1 & 38.5,77,115.5,154,192.5 \\ 
\hline  \hline
 LP2 & Al & 1.5 & 1.7, 2.0, 2.5, 3.0, 4.0, 5.0 & PDMS & -- & $540$ \\\hline
 LP2      &  Al& 0.8, 1.5, 3.0 & 2 & PDMS & -- & $540$ \\ \hline
   &        &1.5 & 2 &  Triton X-305 &   -- & 540 \\
   &        &1.5 & 2 &  SDS &  -- &  540 \\
LP2 &  Al   &1.5 & 2 & oleic acid & 22 &  540 \\
   &        &1.5 & 2 & NBD-PC & 4.66 & 540 \\
   &        &1.5 & 2 &  PDMS &   -- & 540 \\ \hline
\end{tabular}
\end{small}
\caption{{\bf Summary of experiments (all rows)}.  Simulations were performed
with the parameters in the first row. Experiments with
fluorescence imaging (FI) use a silicon wafer (Si) to line the bottom of
the 14.6~cm radius aluminum well. Experiments which only use laser
profilometry (LP) use an anodized 11.1~cm radius aluminum well (Al). Both
lasers are from LaserGlow, with a
250~$\mu$m beam thickness; laser 1 is $20$~mW, $532$~nm, fan
angle $100^\circ$;
laser 2 is $5$~mW, $635$~nm, fan angle $30^\circ$.
Initial concentration $\Gamma_0$ is not applicable for soluble surfactants. The
volumes reported for NBD-PC and oleic acid are that of the surfactant-solvent
solution (see Table~\ref{t:Surfs}); the volatile
solvent evaporates before the experiment begins.
\S\ref{s:growth} and \S\ref{s:decay} report dynamics from only the FI \& LP1 experiments; 
measurements of the distension size are reported for all 5 types of experiments in \S\ref{s:hmax}.}
\label{t:Exper}
\end{table}

Each experiment consists of choosing a particular geometry (retaining
ring radius $R$ and initial glycerin thickness $H$), and a volume $V$
of solvent-dispersed surfactant deposited outside the retaining ring on the surface of the glycerin. These initial conditions are
summarized in Table~\ref{t:Exper}, forming four sets of controlled experiments.
It is helpful to consider several dimensionless quantities which characterize the experiments; we 
calculate these based on anhydrous glycerin at $20$ $^\circ$C.
The Reynolds number $S H^3 \rho / \nu^2 R^2$ (where $S \equiv \sigma_0 - \sigma_s$ is the maximum
reduction in surface tension due to the presence of a particular surfactant) is ${\cal O}(10^{-5})$.
The P\'eclet number $S H / \nu D$ expresses the relative importance of advection with respect to diffusion.
The surfactant diffusivity $D$ of similar surfactants has been reported in the literature to be as low as $10^{-10}$~cm$^2$/s and as high as $10^{-4}$~cm$^2$/s \citep{Agrawal1987}. As a conservative estimate, we take the highest value $D =10^{-4}$~cm$^2$/s \citep{Sakata1969}, for which the P\'eclet number is ${\cal O}(10^3)$.
The Bond number $\rho g R^2 / \sigma_0$, which expresses the relative importance of gravitational forces with respect to capillary forces, is ${\cal O}(10^2)$.
The Ohnesorge number $\sqrt{\nu^2 / \rho S R}$, which expresses the relative importance of viscous stresses with respect to Marangoni stresses, is ${\cal O}(1)$.
The Galilei number $\rho^2 g H^2 R / \nu^2$, which expresses the relative importance of gravitational forces with respect to viscous stresses, is ${\cal O}(10^{-1})$.


Two of the surfactants (NBD-PC and oleic acid) are deposited in solution with a
solvent (chloroform) that quickly evaporates.  This technique
allows us to achieve uniform surfactant concentrations near or below the critical monolayer concentration $\Gamma_c$.  For these surfactants, we can calculate the initial surfactant concentration $\Gamma_0 = \frac{V C}{A}$ where $C$ is the mass of surfactant per unit volume solution and $A$ is the surface area initially covered by surfactant ($A = \pi (r_\mathrm{well}^2 - R^2)$).  We usually express $\Gamma_0$ as a fraction of $\Gamma_c$ and achieve different values of $\Gamma_0$ by depositing different volumes of the solvent-dispersed surfactant solution $V$.

Experiments involving surface tension are quite
sensitive to the preparation protocol. Therefore, we initially clean all
parts using a chemically-appropriate method: detergent (aluminum well), Contrad
70 (glass syringe, stainless-steel retaining ring), and oxygen plasma (silicon
wafer). As a final step, we rinse all parts with 18.2 M$\Omega$ deionized
water and dry with nitrogen gas. In each experiment, we fill the well with 
glycerin measured in a glass syringe, and allow the fluid to
settle for 2 hours to reach a flat state. The retaining ring is lowered by a
nylon line until it just touches the surface of the glycerin, thus dividing the
fluid surface into an inner and outer region. With a micro-pipette, we deposit
the surfactant (or surfactant solution) in multiple drops in the outer region.
A waiting period of 30~min (NBD-PC) or 6~min (all others) is sufficient to
allow the spatial distribution to homogenize and the solvent to evaporate
for all but the lowest concentration.
For the NBD-PC experiments, the initial conditions are readily
described by the concentration $\Gamma_0$, reported as a fraction of the
critical monolayer concentration $\Gamma_c = 0.3 \frac{\mu\mathrm{g}}{\mathrm{cm}^2}$ \citep{Bull1999,Fallest2010a}.
In the case of the bulk
surfactants, the values of $\Gamma_0$ are above $\Gamma_c$ but this value is
nonetheless provided for comparison.  Finally, the ring is lifted by a motor at
a rate of $100 \, \mu$m/s.  
The partially wet stainless-steel ring, while being lifted, draws up a meniscus of glycerin.
Observations begin when the ring separates from the meniscus and is removed from the field of view.
A visual inspection of the ring after it detaches reveals a barely-visible layer of glycerin on the bottom of the ring. The main influence of the wettability of the ring is to set the size of the meniscus; the possible effects of this meniscus will be discussed below.

\begin{figure}
\centering
\includegraphics[width=0.7\linewidth]{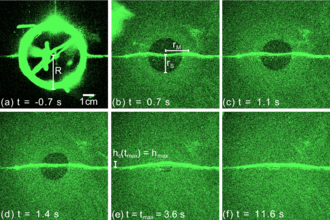}
\caption{[Color Online] {\bf Self-healing of surfactant layer.}
Sample fluorescence imaging (FI) and laser profilometry (LP)
images for $\Gamma_0 = 0.8 \Gamma_c$, $H = 0.7$~mm, and $R=3$~cm. The bright
horizontal
laser sheet reflection measures the height profile $h(r,t)$, and the
background intensity measures the surfactant concentration $\Gamma(r,t)$.
The time series of a typical experiment starts when the retaining ring is
lifted ($a$), after which the capillary ridge travels inward and forms the
vertical distension as the central surfactant-free (dark) region shrinks
($b-d$). At time
$t_\mathrm{max}$, the distension reaches its largest size ($e$) and then decays ($f$).  The position of the maximum of the Marangoni ridge is denoted by $r_M$ and the position of the leading edge of the surfactant is denoted by $r_S$.
}
\label{f:ExpSamples}
\end{figure}

To measure the height profile $h(r,t)$ of the glycerin fluid layer, we use laser
profilometry (LP). The setup consists of a laser sheet generator
centered
along the diameter of the ring, with an incident angle $\approx 20^\circ$.
When the fluid surface is deformed, the laser sheet is deflected by an amount
proportional to the change in the fluid thickness. A camera, positioned
directly
above the experiment, records the reflection of the laser sheet due to both
the top
and bottom glycerin interfaces. In the experiments in which the aluminum well serves
as the bottom interface, the rough surface causes both reflections to appear as
a single profile. In the experiments in which a 8'' silicon wafer serves as the
bottom interface (the FI/LP experiments), profiles
from multiple reflections are visible; an example time series is shown in
Fig.~\ref{f:ExpSamples}. We calibrate the proportionality of the
deflection and the fluid height using flat glycerin layers of known
thickness.
To interpret the profile from the raw images, we trace the center of the
single profile (LP data set), or the top edge of the uppermost (thinnest)
profile (FI/LP data sets).  Each column provides a single value in $h(r)$ to
within one pixel ($0.037$~mm);
individual columns (and, infrequently, images) are rejected from the data set if
they are statistical outliers arising from image imperfections.

To measure the surfactant concentration profile $\Gamma(r, t)$ in experiments
using NBD-PC surfactant, we use fluorescence imaging (FI).  In experiments where
we use this technique, the basic apparatus is modified in several ways.
Eight blue LEDs (1.5~W, $467$~nm from Visual Communications Company, Inc.) are arranged in an evenly spaced
circle around the edge of the aluminum well so
that direct rays of light are reflected away from the camera by the silicon
wafer placed in the bottom of the aluminum well. These LEDs are
close to the $464$~nm absorption peak of the NBD fluorophore. Imaging
is performed with a cooled $14-$bit Andor Luca$-$R camera with $1004 \times
1002$
resolution. The camera is fitted with a Newport bandpass
filter at $(530 \pm 10)$~nm to preferentially collect photons emitted near the
$531$~nm emission peak of the NBD fluorophore. The exposure time for each image
is $1/4$~s with a frame rate of $2-3$ Hz, which allows for a high enough signal-to-noise ratio to image
concentrations down to $\lesssim 0.05 \Gamma_c$.

\begin{SCfigure}
\includegraphics[width=0.6\linewidth]{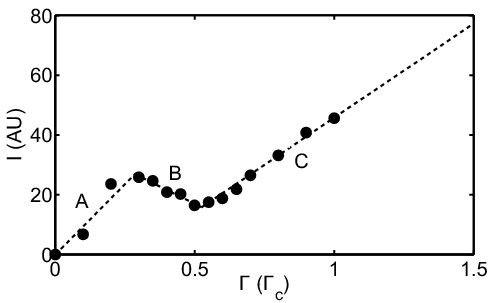}
\caption{{\bf Calibration data relating surfactant concentration
$\Gamma$ to fluorescence intensity $I$.} Because the curve is
non-monotonic, we use a piecewise linear fit (regions A, B, C) to the
empirical data. Error bars (standard deviation of the mean intensity) are smaller than the points.}
\label{f:IGCal}
\end{SCfigure}

We calibrate the relationship between image intensity and surfactant
concentration $\Gamma$ by depositing a known concentration of NBD-PC on a flat
glycerin surface and recording the average brightness within a
$10 \times 10$~cm$^2$ region at the center of the image. At
the lowest concentrations, ($< 0.3 \Gamma_c$), the surfactant remains
inhomogeneously distributed despite waiting 2 hours after deposition. For these
experiments, we illuminate the experiment with
only the blue LEDs. In addition, the calibration
accounts for a temperature-dependent offset value which is manifest as a slow
drift in the average image intensity.  As is expected for molecular
fluorescence, the
relationship is non-monotonic due to fluorescence resonance energy transfer
(FRET) effects \cite{Shrive1995}. The resulting calibration is shown
in Fig.~\ref{f:IGCal}.

To calculate $\Gamma(r)$ from the images, we first crop and threshold the image to remove
imperfections, and mask the region containing the laser profile. As a
consequence, we do not report FI data for the central $8$~mm of the
image. Using the remaining values, we azimuthally average the image intensity
in 2 pixel wide bins. To account for the slow temperature drift, we set
$\Gamma \approx \Gamma_0$ for large $r$ and, if needed, $\Gamma \approx 0$ at the center. Because
$I(\Gamma)$ is non-monotonic, we must develop an inversion
procedure to measure $\Gamma(I(r))$. We start from a piecewise-linear fit to
$I(\Gamma)$ (see Fig.~\ref{f:IGCal}),
and use a combination of interpolation and continuation methods to perform the
inversion.  For regions where $\Gamma(I)$ is ambiguous, 
we assume continuity of $\Gamma(I(r))$ and extrapolate from
unambiguous regions by identifying the correct piecewise regime (A, B, C from
Fig.~\ref{f:IGCal}) to use. This method even works when $I(r)$ crosses through
all 3 regimes within a few pixels, as occurs at the leading edge of the
inward-spreading surfactant, by relying on extrapolation from both sides. 
 Note that $\Gamma(r)$ curves can rise above the
initial $\Gamma_0$ value by drawing material from other regions.

\subsection{Model \label{s:model}}

The classic Gaver-Grotberg model \cite{Gaver1990} for the spreading of
insoluble surfactant on a thin liquid film is based upon low Reynolds number flow in a small aspect ratio system.
Marangoni stresses at the fluid surface drive flow in the bulk. The velocity of the flow is 
determined by the no-slip boundary condition at the container bottom. 
The normal and tangential stresses, including
gravity (via hydrostatic pressure) and capillarity (via normal surface stress condition), 
balance at the fluid surface.
The resulting system of nonlinear partial differential equations is
    \begin{equation}\label{e:film}
    \tilde{h}_{\tilde{t}} + \nabla \cdot \left( \frac{1}{2} \tilde{h}^2 \nabla \tilde{\sigma} \right) =
    \beta \nabla \cdot \left( \frac{1}{3} \tilde{h}^3 \nabla \tilde{h} \right) -
    \kappa \nabla \cdot \left( \frac{1}{3} \tilde{h}^3 \nabla \nabla^2 \tilde{h} \right)
    \end{equation}

    \begin{equation}\label{e:surfactant}
    \tilde{\Gamma}_{\tilde{t}} + \nabla \cdot \left( \tilde{h} \tilde{\Gamma} \nabla \tilde{\sigma} \right) =
    \beta \nabla \cdot \left( \frac{1}{2} \tilde{h}^2 \tilde{\Gamma} \nabla \tilde{h} \right) -
    \kappa \nabla \cdot \left( \frac{1}{2} \tilde{h}^2 \tilde{\Gamma} \nabla \nabla^2 \tilde{h} \right) +
    \delta \nabla^2 \tilde{\Gamma}.
    \end{equation}
where $\tilde{h}({\bf \tilde{r}},\tilde{t})$ is the fluid height and $\tilde{\Gamma}({\bf \tilde{r}},\tilde{t})$ is the surfactant concentration.
For clarity, all non-dimensional variables are denoted with a tilde to distinguish them from their dimensional analogues (e.g. $\tilde{t}$ is a non-dimensional quantity while $t$ has dimensions of seconds).
In Eqn.~\ref{e:film}, gravity smooths the height profile, driving the surface to be flat while capillarity affects curvature in the fluid surface \cite{Peterson2010}.
For simplicity, capillarity is assumed to be a property of the fluid and independent of the surfactant-fluid interaction. 
The fluid height evolution equation is derived from fluid incompressibility where the convective derivative uses the depth-averaged velocity.
The surfactant evolution equation is derived from an advection-diffusion equation in which the surface convective derivative contains the fluid surface velocity and the surfactant diffuses on the surface.  The diffusion smooths the surfactant profile, slowly causing the surfactant to reach a uniform coverage independent of Marangoni stresses.
In both equations, the $\nabla \tilde{\sigma}$ term 
($\nabla = \partial_{\tilde{x}}\hat{x} + \partial_{\tilde{y}}\hat{y}$)
describes the effect of gradients in surfactant
concentration.

\begin{SCfigure}
\includegraphics[width=0.4\linewidth]{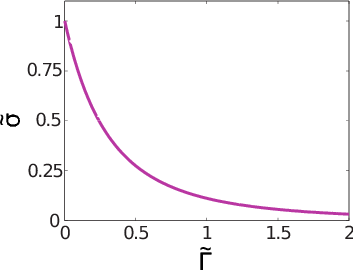}
\caption{{\bf The equation of state used in the simulations (Eqn.~\ref{e:EOS}).}  The equation of state relates the nondimensional surface tension $\tilde{\sigma}$ to the nondimensional surfactant concentration $\tilde{\Gamma}$.}
\label{f:EOS}
\end{SCfigure}

The relationship between surface tension and
surfactant concentration is provided by an equation of state. We use
\begin{equation}\label{e:EOS}
\tilde{\sigma}(\tilde{\Gamma})=\left(1 + \eta \tilde{\Gamma} \right)^{-3},
\end{equation}
(see Fig.~\ref{f:EOS})
which was proposed by \citet{Borgas-1988-MFT} as an alternative to the linear
equation of state because it can model concentrations beyond a monolayer of
surfactant \cite{Swanson2013}. 
Empirical measurements of the equation of state of lipids typically have
a regime where the surface tension remains constant for concentrations greater than $\Gamma_c$, 
above which the  monolayer becomes sufficiently close-packed that no additional molecules can fit on the fluid surface \citep{Kaganer1999}. For $\Gamma > \Gamma_c$, three-dimensional structures are created, 
but because surfactant molecules only lower the surface tension when they are in contact with the surface, the surface tension does not change further. In Eqn.~\ref{e:EOS}, 
the material parameter $\eta \equiv \frac{\sigma_s}{\sigma_0 - \sigma_s}$
is determined from both the surface tension $\sigma_0$ of the surfactant-free glycerin and
that of the surfactant-contaminated fluid ($\sigma_s$) when $\Gamma = \Gamma_c$.
Based on material parameters for NBD-PC on glycerin, we choose $\eta = 1.2562$.
For the NBD-PC experiments,  $\sigma_s$ is set by the empirically-determined
minimum surface tension \cite{Bull1999} which is reached at or
above the critical monolayer concentration, $\Gamma_c$.
For a monolayer of molecules, this molecular thickness is a negligible contribution to $\tilde{h}(\tilde{r})$.

There are three non-dimensionalized coefficients in the model, each of which
controls the magnitude of a particular term. The second-order $\beta$ terms incorporate the effect of gravity, while the fourth-order $\kappa$ terms model the effect of capillarity on the system.
The surfactant equation also has a $\delta$ term representing surfactant diffusion on the surface of the film.
We compute these three coefficients according to
\begin{equation}
 \beta \equiv \frac{\rho g H^2}{S},\qquad
 \kappa \equiv \frac{\sigma_0 H^2}{S L^2},\qquad
 \delta \equiv \frac{\nu D}{SH}\label{nondim}
\end{equation}
using physical values from Table~\ref{t:Surfs} and the
dimensions of our experiments. Here, $\rho$ is the fluid density, $\nu$ is
the fluid viscosity, $g$ is the acceleration due to gravity, $L$ is the
characteristic length (taken to be the ring radius $R$), and $H$ is the
characteristic fluid height (taken to be the initial fluid height). The validity
of the lubrication approximation, an important assumption of the model, is
limited by how thin we can make the glycerin layer before it dewets during
surfactant deposition. Our thinnest stable films are $H=0.7$~mm
thick. For the largest retaining ring ($R=3$~cm), this provides an aspect ratio
of $\epsilon = 0.02$. The parameter $S$ 
is determined by the range of surface tension values accessible to the materials in the system.
From the diffusivity ($D = 10^{-4}$~cm$^2$/s),
we calculate the time scale for purely diffusive self-healing of our monolayer
films as $R^2/D = 10^4 - 10^5$~s (2 to 25 hours, depending on the size of the ring).  
Using these values, together with the parameters in Table~\ref{t:Exper}, we calculate 
$\beta = 2.13612 \times 10^{-1}$,
$\kappa = 1.22839 \times 10^{-3}$, and $\delta = 7.17844 \times 10^{-4}$.

We calculate numerical solutions using an open source numerical solver \citep{ClaridgeLevyWong} designed to solve 
coupled hyperbolic-parabolic nonlinear partial differential equations, such as Eqn.~\ref{e:film} and Eqn.~\ref{e:surfactant}.  The package employs a finite volume scheme for nonlinear systems of PDEs up to fourth-order without restrictions on
boundary conditions, minimizes per-problem code-development, and enables rigorous,
automated convergence testing on problems with analytical solutions. 
We use a rectangular grid 
so as not to impose a symmetric solution via the solver \citep{SIURO}, and find that the solutions are axisymmetric.  Without loss of generality (and to facilitate direct comparison with the azimuthally averaged experimental data), we present cross-sections of the redimensionalized height and surfactant profiles in the $y = 0$ plane (coplanar with $\theta = 0$ plane).  
Additionally, the abscissa of the height and surfactant profile plots in \S\ref{s:results} is labeled $r$ instead of $\tilde{x}$.
The distance between cells is 0.035 dimensionless spatial units; 
we compute on the domain $[-3.4825, 3.4825]$ 
(in dimensional units: $[-10.4475\,\mathrm{cm}, 10.4475\,\mathrm{cm}]$)
using Dirichlet boundary conditions.  Convergence of the solver's solutions for similar problems
has been shown in previous work \citep{SIURO,ClaridgeLevyWong}.  We do not
consider reflections from the edges of the well, which is appropriate
for the inward spreading case.

\begin{figure}
\includegraphics[width=0.8\linewidth]{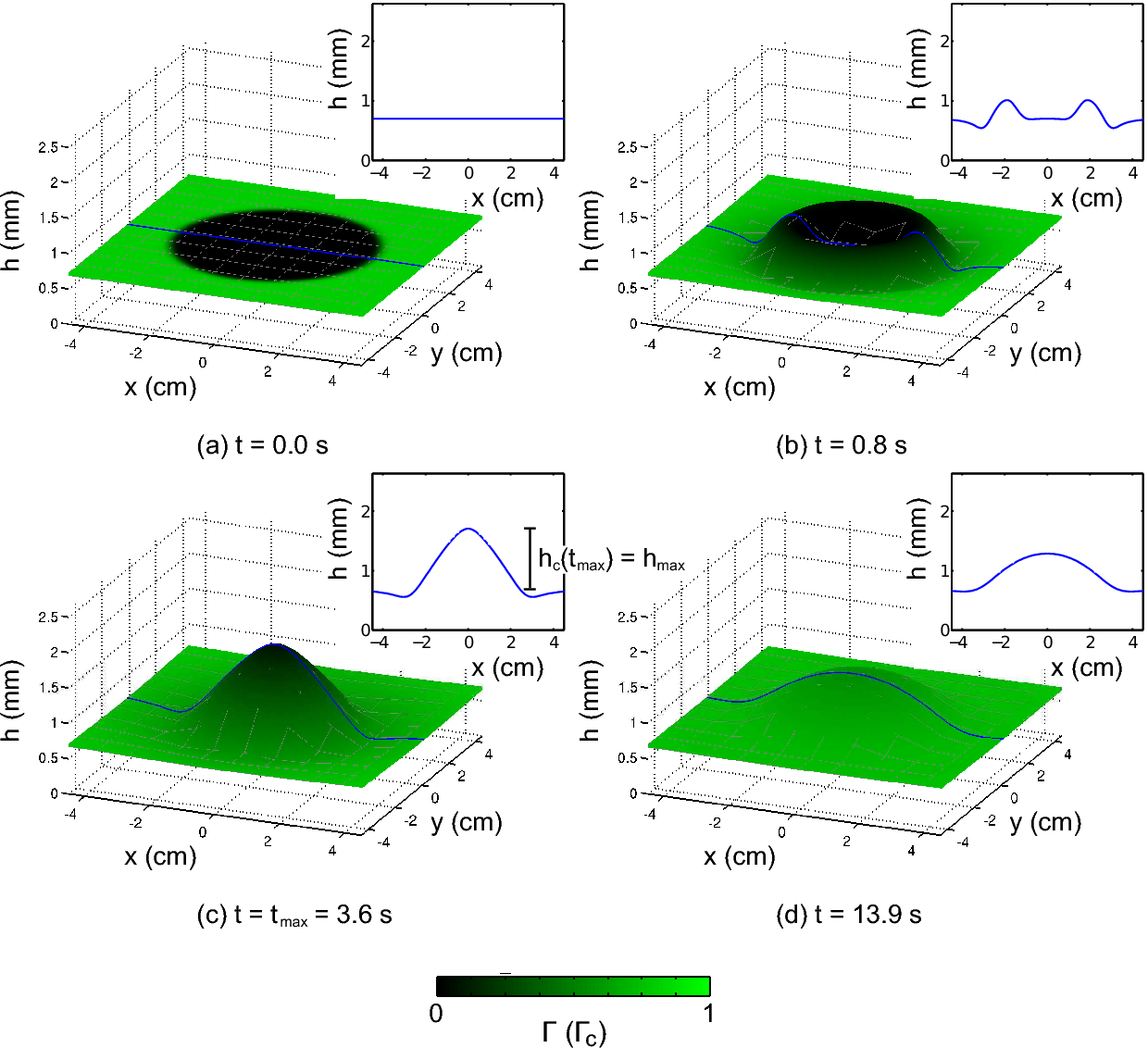}
\caption{[Color Online] {\bf Typical numerical simulations with growth and decay of fluid and self-healing of surfactant.}  Redimensionalized numerical simulations of Eqn.~\ref{e:film} and
Eqn.~\ref{e:surfactant} using $\Gamma_0 = 0.8 \Gamma_c$, $H = 0.7$~mm, and
$R=3$~cm (the same parameters as in Fig.~\ref{f:ExpSamples}). In each panel,
the surface represents $h(x, y, t)$ and the color of the surface ranges
from $\Gamma = 0 \Gamma_c$ (black) to $\Gamma = 1 \Gamma_c$
(green). The insets highlight the shape of $h(y)$ cut through the center of the
surface plots. The simulation starts from
($a$) the initial film height and surfactant concentration, and the resulting
dynamics are ($b$) the formation of the distension as the annular Marangoni
ridge coalesces toward the center of the region.  At time
$t_\mathrm{max}$, the distension reaches its largest size ($c$) and then decays ($d$).
Times have been redimensionalized by $T / \alpha_{0.8}$, as will be discussed
in \S\ref{s:results}, to permit a direct comparison with the experiments.}
\label{f:SimSamples}
\end{figure}

To model the initial condition of the physical experiment, we use a uniform
initial film height of $\tilde{h}=1$ and an initial surfactant concentration of
\begin{equation}
\tilde{\Gamma}(\tilde{r};\tilde{t}=0) = \tilde{\Gamma}_0 \frac{1}{2} \left[1 - \tanh\left( 20 \left(1 - \tilde{r} \right) \right) \right],
\end{equation}
where $\tilde{r}$ is the radial coordinate. Sample solutions representing the resulting
inward-spreading dynamics for NBD-PC are shown in Fig.~\ref{f:SimSamples}. Each
panel represents both  $h(r,t)$ and $\Gamma(r,t)$ (redimensionalized for comparison) at the same instant.

When making comparisons between experimental results and numerical solutions of
this model,  we reintroduce dimensions into the numerical results using $h = \tilde{h} \, H,$ 
$\Gamma = \tilde{\Gamma} \, \Gamma_c$ and $\Gamma_0 = \tilde{\Gamma}_0 \, \Gamma_c$. The timescale is expected \cite{Gaver1990} to
be similarly redimensionalized according to time $t = \tilde{t} \, T$, where
\begin{equation}
  T =\frac{\nu L^2}{SH}.
\label{eqn:timescale}
\end{equation}
However, we observe that the simulated fluid height profile and surfactant distribution both evolve more slowly than the experiment by a factor of approximately 10.  This discrepancy in the speed of the dynamics indicates that the redimensionalization timescale is incorrect.
One possible explanation is a decrease in the viscosity of the glycerin layer, as described in
\S\ref{s:exper}. To account for this discrepancy in timescale, we introduce an empirical factor $\alpha$ into the redimensionalization: $t = \tilde{t} \frac{T}{\alpha}$. Effectively, this speeds up the simulations by a factor of $\alpha$. The choice of $\alpha$ will be described in \S\ref{s:results}, and all simulation results will have the times redimensionalized in this way.

\section{Results \label{s:results}}

In both experiments and simulations, we observe a self-healing phenomenon,
whereby a surfactant on a contaminated surface spreads inward, covering the
initially surfactant-free region.  The surfactant-depleted region can
persist for several minutes, and the closing process is accompanied by a
corresponding rise in the fluid level at the center of the region. In this
section, we characterize the
dynamics of this process and evaluate the efficacy of the model in capturing the
phenomenon.  For simplicity, we report all variables in their dimensionalized form.

Because our experiments capture the spatiotemporal evolution of the
surfactant concentration profile $\Gamma(r, \theta, t)$ and height profile $h(r, t)$
for different initial
concentrations of surfactant, we can compare the simulation results to the
experiments outlined in Table~\ref{t:Exper}.  To characterize the dynamics of
the growth of the distension, we  consider $h_c(t)$, the height of the distension
at the center of the domain, as well as $h_\mathrm{max}=h_c(t_\mathrm{max})$, the maximum height attained.
Note that in the FI/LP experiments (and their corresponding
simulations), only monolayer concentrations of NBD-PC surfactant are explored,
while the LP experiments include bulk surfactants. In the LP experiments, we
can nonetheless characterize the dependence of $h_\mathrm{max}$ for different initial
fluid heights, ring sizes, and types of surfactant.  We do not report
simulations for the LP (bulk surfactant) experiments, since they are far beyond
the monolayer regime and therefore violate the assumptions of the
model.

Fig.~\ref{f:ExpSamples} shows a prototypical inward spreading (FI/LP)
experiment, visualized from above. The first image (a) shows the ring before
it is lifted; light reflected from the ring obscures the lipid fluorescence.
In (b-f) the speckled green outer region indicates the presence of surfactant,
and the black central region is initially surfactant-free.
The bright green curve through the center of the images is the intersection of
the laser sheet with the surface of the fluid, which provides a profile of the
fluid height.  A simulation using the same material parameters and initial
geometry is shown in Fig.~\ref{f:SimSamples}, where the green shading represents
the surfactant concentration, and the blue curve in the insets provide a view
analogous to the green laser profile.


\begin{figure}
\centerline{\includegraphics[width=0.7\linewidth]{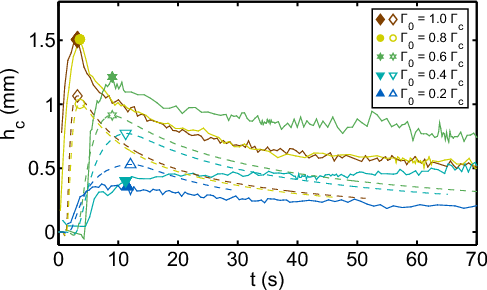}}
\caption{[Color Online] {\bf Distension growth and decay.}
Dynamics are captured by tracking height $h_c$ at
the center of the system. Experiments (filled symbols) and redimensionalized simulations (open
symbols) at five different initial surfactant concentrations.
As discussed in \S\ref{s:model},} simulation times have been
redimensionalized and then further reduced by a factor
${\alpha}_{0.2} = 17.2 {\genfrac{}{}{0pt}{}{+12}{-0.0}}$,
${\alpha}_{0.4} = 7.81 \pm 2$,
${\alpha}_{0.6} = 6.07 \pm 1$,
${\alpha}_{0.8} = 11.0 \pm 1$, and 
${\alpha}_{1.0} = 9.91 \pm 2$, respectively, where the $\pm$ values represent
the uncertainty in $\alpha$.
\label{f:hct}
\end{figure}

In experiments and simulations, similar dynamics are observed. First, a fluid distension forms as
the annular Marangoni ridge coalesces in the center of the surfactant-free
region. The distension reaches a maximum height $h_\mathrm{max}$ at time
$t_\mathrm{max}$, after which the distension decays. 
To quantitatively compare experimental data with simulations, we 
redimensionalize the simulation data as described in
\S\ref{s:model}.  As illustrated in Fig.~\ref{f:hct}, we empirically determine the
best redimensionalization timescale $T/\alpha$ by comparing
the dynamics at the center of the distension, $h_c(t)$. For each value of
$\Gamma_0$, we choose $\alpha$ so that both the rise time and the observed
$t_\mathrm{max}$ approximately coincide. In all cases, we find that
the simulation timescale
needs to be decreased by a factor of approximately 10. The resulting
$\Gamma_0$-dependent values of $\alpha$ are provided in the caption to
Fig.~\ref{f:hct} and used throughout the remainder of the paper.

Fig.~\ref{f:hct} characterizes the dynamics of the distension growth and
decay at the center of the system, comparing $h_c(t)$ for five
different initial surfactant
concentrations drawn from the FI/LP experiments. We find that the simulation is
able to semi-quantitatively capture a number of key features observed in the
experiment, beginning with the observation that the distension is
approximately 1~mm high. In both simulations and experiments, we observe that
the maximum value of $h_c$ increases with surfactant concentration (and thus
increased surface tension contrast). This trend will be explored quantitatively
in \S\ref{s:hmax}.  In all cases, the growth process is faster than the decay
process. In addition, both simulations
and experiments show a transition in the sharpness of the growth/decay dynamics ($h_c$),
with a more pronounced peak present for larger values of $\Gamma_0$.
Intriguingly, the $1.0$ and $0.8, \Gamma_c$ curves are almost coincident with
each other in both the experiments and simulations. One key disagreement between experiment in simulations  is that for
the $1.0, 0.8$ and $0.6 \, \Gamma_c$ experiments, $h_\mathrm{max}$ is 0.3 to 0.4~mm
taller in the experiments than in the simulations.  However, we note that this
difference corresponds to the height of the meniscus formed when the ring
lifts from the fluid surface. Second, we observe that for experiments starting
from $0.2$ and $0.4 \, \Gamma_c$ initial conditions, the meniscus is large
enough to obscure the observation of the distension, whereas in the simulations, the
distension is clearly present.

The growth and decay dynamics at higher concentrations can be understood as
arising from surface
tension gradients. In the growth phase, fluid with higher surface tension
(fluid initially surfactant-free inside the ring) pulls the
surfactant-rich low surface tension fluid inward.  The fluid advects the
surfactant and heals (closes)
the surfactant layer hole while horizontal gradients in the velocity
field result in an annular Marangoni ridge.  This ridge moves
toward the center, and coalesces into the central distension. In the decay
phase, the distension decreases in height until the surface has uniform height
while the surfactant concentration homogenizes. In the sections that follow, we
quantitatively examine the spatial distributions of surfactant which underlie
these dynamics of these two phases.

\subsection{Distension Growth \label{s:growth}}

\begin{figure}
\centerline{
\includegraphics[width=0.8\linewidth]{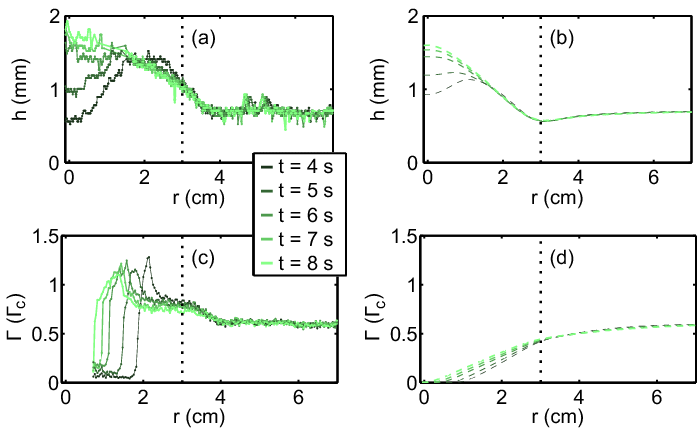}
}
\caption{[Color Online]  {\bf Distension growth dynamics.} 
($a$,$c$) Experiments (solid) and
($b$,$d$) redimensionalized simulations (dashed) of distension growth
dynamics for initial conditions $\Gamma_0 = 0.6 \, \Gamma_c$, $R=3$~cm, and $H=0.7$~mm.
Height profile $h(r,t)$ ($a$,$b$) and surfactant concentration profile $\Gamma(r,t)$ ($c$,$d$) at
representative times during growth.
Vertical dotted line indicates the initial ring location at $r=R$.
Note that gradients in the surface tension field are not significant when $\Gamma > \Gamma_c$.
\label{f:Growth}
}
\end{figure}

During the growth phase, the motion of the fluid both raises a central
distension and advects surfactant inward. 
In Fig.~\ref{f:Growth}, we show an example ($\Gamma_0 = 0.6 \Gamma_c$) of the evolution
of both $h(r,t)$ and $\Gamma(r,t)$ during the growth of the distension. Here,
and also in several plots that follow, the center of the system is at $r=0$,
and therefore inward spreading corresponds to motion to the left. In addition,
note that at a given time, $\Gamma(r;t)$ represents the {\itshape concentration} of a molecular
monolayer, and that the thickness of this layer is many orders of magnitude
smaller than $h(r;t)$. In panels $a$ and $b$, we
observe that the simulations semi-quantitatively capture the observed dynamics
of $h(r,t)$.

\begin{SCfigure}
\includegraphics[width=0.6\linewidth]{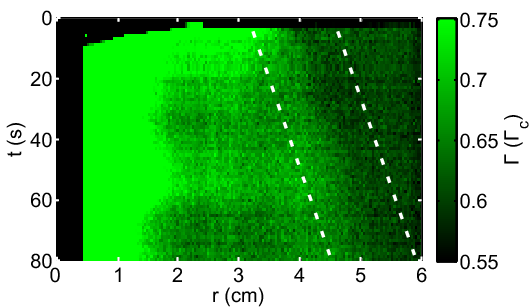}
\caption{[Color Online]
{\bf Kymograph of the surfactant concentration profile for the 0.6 $\Gamma_c$ experiment.}  Surfactant propagates outward in the region demarcated by the two white dashed
lines. The color scale is truncated at  $0.55\Gamma_c$  and $0.75\Gamma_c$ in
order to isolate the region of interest. For reference, the maximum surfactant
concentration within the green colored saturated region is $1.4$ $\Gamma_c$.}
\label{f:SurfSpaceTime}
\end{SCfigure}

In contrast, we find that the simulated $\Gamma(r,t)$ (panel $d$) takes a quite
different shape from what we have observed in experiments (panel $c$); similar
disagreement has been found for the case of droplet-spreading
\citep{Fallest2010a,Swanson2013}.
First, we observe that the experiments exhibit a sharp (1 mm) interface at the
leading edge of the advancing front of surfactant (location $r_S$). In contrast,
the simulations show transition region with a width similar to retaining ring
radius $R=3$~cm.
Second, we observe a sharp peak in the surfactant concentration directly behind
$r_S$ not present in the simulations. Visual inspection of the original images suggests that this feature
arises from the lift-off of the retaining ring, and it may be that these
dynamics cause the formation of a bilayer or other condensed phase
\citep{Kaganer1999}. 
On each side of this peak there is a gradient in $\Gamma$; the opposite signs
indicate that the surfactant should spread in both directions. Indeed, we see
both the leading edge and the peak travel toward the center of the experiment
(self-healing). Simultaneously, the shallower gradient that trails the peak
forms an enhanced surfactant plateau that moves outward. This feature is
visible in Fig.~\ref{f:SurfSpaceTime}.  If this outward
propagation of surfactant is advection, then 
the fluid must bear outward surface currents, an effect absent in the simulation but 
predictable from the balance of Marangoni stress and viscous shear stress
present in lubrication theory.   
Both features -- the peak in $\Gamma$ and the trailing plateau --
are notably absent from the simulations. A leading plateau was previously
observed in outward-spreading experiments \citep{Fallest2010a, Swanson2013}, and
models have so far been unable to capture either feature \citep{Swanson2013}.
Its absence is likely due to both a failure to account for surfactant build up
on the meniscus and the choice of the equation of state (Eq.~\ref{e:EOS}), which does
not account for any condensed phases.  Were
such structures present, it is possible that they are advected by the underlying fluid
velocity. In spite of these difficulties, the simulated $\Gamma(r,t)$ is able to
capture surfactant self-healing.

\begin{figure}
\includegraphics[width=0.8\linewidth]{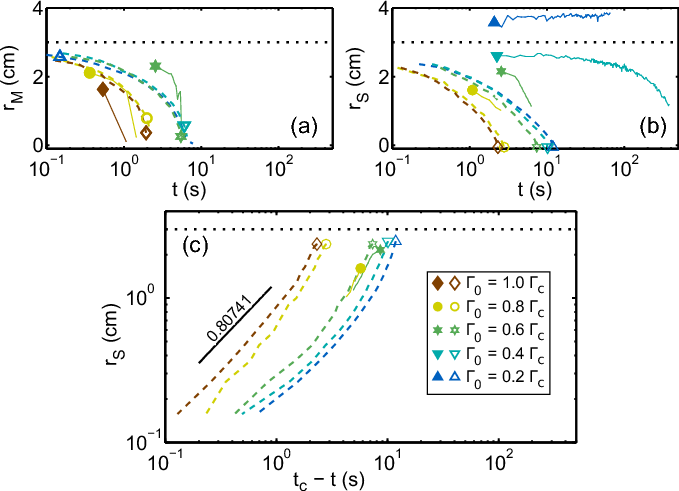}
\caption{[Color Online] {\bf Maragoni ridge and surfactant leading edge dynamics.}
($a$) The peak of the Marangoni ridge ($r_M$) and
($b$) the inward motion of the leading edge of the surfactant ($r_S$) at five
different initial surfactant concentrations, for both experiments (solid
lines) and redimensionalized simulations (dashed lines). For the experiments, all locations are
determined by visual inspection of the $h(r,t)$ and $\Gamma(r,t)$ profiles.
In the simulations, $r_S$ is the first point
where the surfactant concentration rises above $0.001$ $\Gamma_c$; choosing a
lower threshold does not significantly affect the results.
In all cases, $R=3.0$~cm and $H=0.7$~mm; the horizontal dotted line indicates the
initial ring location at $r=R$.
The closing time $t_c$ is defined as time at which $r_S = 0$; in the
experiments, this is found by visual inspection of the images.
($c$) The position of the leading edge of the surfactant (same data as $a$),
but plotted as a function of ($t_c - t$) to permit comparison
to the asymptotic solution from \citet{Jensen1994} (black line).
The symbols identify the initial conditions used for each curve.}
\label{f:rS-growth}
\end{figure}

During the growth phase, we quantify the spreading dynamics by measuring the
location of both the peak of the Marangoni ridge ($r_M$) and the leading
edge of the surfactant ($r_S$), with respect to the center at $r=0$.
These dynamics are shown in Fig.~\ref{f:rS-growth}$a$,$b$. No data is shown for
$r_M$ at $\Gamma_0 = 0.2 \, \Gamma_c$ and $0.4 \, \Gamma_c$ because the
Marangoni ridge, if present, is obscured by the meniscus. No data is shown for
$r_S$ at  $\Gamma_0 = 1.0 \Gamma_c$ (and limited data at $\Gamma_0 =
0.8 \Gamma_c$ and $0.6 \Gamma_c$) because the dynamics were faster than could be
captured by the optics. We observe that $r_S$ and $r_M$ approximately coincide
in all cases,
 but that self-healing (if it is observed), occurs after the
fully-formed distension at $t_\mathrm{max}$. An example of these dynamics are shown in 
Fig.~\ref{f:Growth} for $\Gamma_0 = 0.6 \, \Gamma_c$. In all cases except for
$\Gamma_0 = 0.2 \Gamma_c$, we observe that
self-healing occurs within $10^2$~s (a few minutes), which is far shorter than
the diffusive timescale of $10^5$~s. Therefore, the self-healing is a
Marangoni-driven
process. As such, the closing dynamics in both simulations and experiments are
faster for larger $\Gamma_0$, as would be expected due to the larger surface
tension gradients. In simulations, unlike experiments, self-healing always
occurs.

Self-healing was predicted by \citet{Jensen1994} to take the asymptotic form
$r_S(t) \propto (t_c-t)^{0.80741}$, where $t_c$ is the time at which $r_S$
reaches zero and the surfactant-free region is closed. 
In Fig.~\ref{f:rS-growth}$c$, we compare this form to our simulations and
experiments; note that here time progresses from right to left. The simulations agree as 
expected because they are solutions of the same equations, 
but with the $\beta$ and $\kappa$ terms included.
For experiments with $\Gamma_0 = 0.6$ or $0.8 \, \Gamma_c$ (the two
experiments with sufficient data available, although less than a decade), we
observe approximate agreement with the predicted exponent.
Additionally, these figures show that 
the choice of $\alpha$ for analyzing $t_\mathrm{max}$ 
(based upon fluid height profile data)
also yields approximately correct values for $t_c$ for the three runs with the highest
initial surfactant concentrations. 
In contrast, $t_c$ for the $0.4 \, \Gamma_c$ experiment is more than an order of magnitude larger than the redimensionalized simulation, and $t_c$ for the 0.2 $\Gamma_c$ experiment could not be measured. This indicates that a different process controls the timescale at low concentrations.

\begin{figure}
\centerline{
\includegraphics[width=0.8\linewidth]{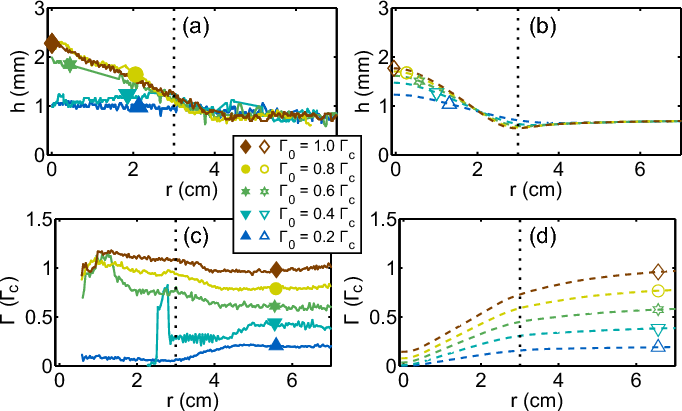}
}
\caption{ {\bf Profiles at $t_\mathrm{max}$ for range of $\Gamma_0$.} ($a$, $b$) Fluid height and ($c$,$d$) surfactant concentration
profiles for FI/LP experiments and redimensionalized simulations at time $t_\mathrm{max}$, starting from
five values of $\Gamma_0$
with $R=3.0$~cm and $H=0.7$~mm. 
($a$, $c$) Experiments and ($b$,$d$) simulations are plotted in solid and dashed lines respectively.
Vertical dotted lines indicate the initial ring location at $r=R$.  From lowest
to highest $\Gamma_0$, $t_\mathrm{max}$ is $12.08$, $11.25$, $9.02$, $3.60$, and $3.21$
(s).
The symbols identify the initial conditions used for each curve.}
\label{f:ProfGamma}
\end{figure}

The growth phase ends when the distension reaches its maximum height
$h_\mathrm{max}$ at time $t_\mathrm{max}$.  In Fig.~\ref{f:ProfGamma}, we compare $h(r,t)$ and
$\Gamma(r,t)$ at this time. We observe semi-quantitative agreement in $h(r,t)$ for
experiments and simulations, but only for $\Gamma_0 = 0.6, 0.8,$ and $1.0 \,
\Gamma_c$. At lower initial concentrations, the formation of a central
distension is suppressed only in the experiments. Once again the surfactant
concentration profiles $\Gamma(r,t)$ show much less agreement between simulations
and experiments.  For all runs, the simulations have self-healed ($\Gamma(0) >
0.0001 \, \Gamma_c$) by the time of $t_\mathrm{max}$ but retain a strong concentration
gradient. In contrast, the experiments self-heal (if at all), after $t_\mathrm{max}$.

\subsection{Distension Decay \label{s:decay}}

\begin{figure}
\centerline{
\includegraphics[width=0.8\linewidth]{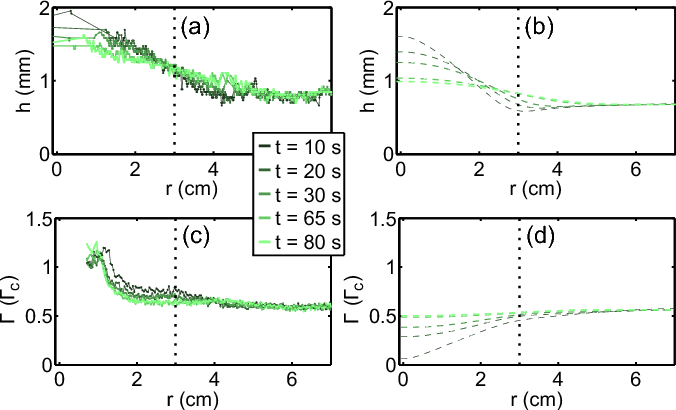}
}
\caption{[Color Online] {\bf Distension decay dynamics.}
($a$,$c$) Experiments (solid) and
($b$,$d$) redimensionalized simulations (dashed) of distension decay
dynamics for initial conditions $\Gamma_0 = 0.6 \, \Gamma_c$, $R=3$~cm, and
$H=0.7$~mm.
Height profile $h(r,t)$ ($a$,$b$) and surfactant concentration profile $\Gamma(r,t)$ ($c$,$d$) at
representative times during decay.
Vertical dotted line indicates the initial ring location at $r=R$.
\label{f:Decay}
}
\end{figure}

During the decay phase of the
distension after $t_\mathrm{max}$ the fluid returns from  $h_\mathrm{max}$ to its original
uniform height and $\Gamma(r,t) = const.$
Fig.~\ref{f:Decay} shows this continuation of the growth dynamics begun
in Fig.~\ref{f:Growth}; here, because the dynamics have
slowed, we examine profiles at longer intervals of time.
Three driving forces are at work in the decay:
gravitational forces (decreasing $h_\mathrm{max}$), capillary forces (decreasing $h_\mathrm{max}$), and
Marangoni forces (equilibrating the surfactant distribution and smoothing $\Gamma(r,t)$).
In the experiment, the sign of $\nabla \Gamma(r,t)$ indicates that although the
surfactant is still being advected inward by the
fluid, it is also spreading outward to equilibrate the
concentration, as is visible in Fig.~\ref{f:SurfSpaceTime}.  
Again, while the simulations show reasonable agreement for $h(r,t)$, they 
fail to capture the observed surfactant concentration profile. In the
simulations, the decay dynamics simply involve the inward motion of the
surfactant as the concentration gradient relaxes monotonically.

\begin{SCfigure}
\includegraphics[width=0.6\linewidth]{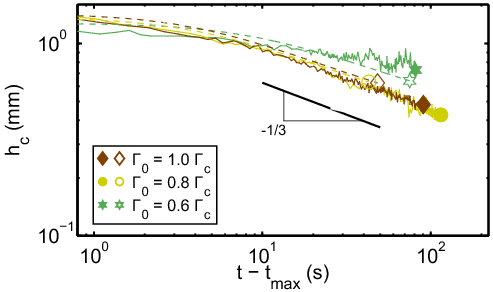}
\caption{{\bf Power law decay near $\Gamma_c.$}  Comparison of experiments (solid lines) and redimensionalized simulations (dashed lines) at three values of
$\Gamma_0$, highlighting the distension decay dynamics using data from 
Fig.~\ref{f:hct}.
The simulations account for the presence of the
meniscus by adding the experimentally-observed meniscus height.
The solid line is a
comparison to a power law decay with exponent $-1/3$, provided for comparison.
The symbols identify the initial conditions used for each curve.}
\label{f:hcDecay}
\end{SCfigure}

To examine the asymptotic behavior of the decay process, we examine the
decreasing distension height $h_c$ as a function of $t-t_\mathrm{max}$. 
Fig.~\ref{f:hcDecay} provides a comparison of these dynamics for
experiments and simulations. In order to make a direct comparison, we
added the experimentally-observed meniscus height ($0.36$~mm) to all values of
$h_c$ in the simulations. For $\Gamma_0 = 0.6, 0.8,$ and $1.0 \, \Gamma_c$, we
find quantitative agreement between experiments and simulations, in spite of
the disagreement in the spatial distribution of surfactant. In the long-time
limit, the asymptotic decay dynamics appear to be governed by
\begin{equation}
 h_c(t) \propto (t-t_\mathrm{max})^{-1/3},
\end{equation}
with a slightly smaller exponent for $\Gamma_0 = 0.6 \, \Gamma_c$.
We exclude the data for $\Gamma_0 = 0.2 $ and $0.4 \, \Gamma_c$ from this
comparison  because the height at the center is changing primarily due to the
effect of the meniscus (rather than the surfactant).

\subsection{Distension Size \label{s:hmax}}

In addition to the FI/LP experiments with NBD-PC as the surfactant, we also
perform experiments with a variety of other common
surfactants, using both monolayer and bulk concentrations. The geometries of
these experiments, and the properties of these surfactants are summarized in
Tables~\ref{t:Surfs} and \ref{t:Exper}. To isolate geometric effects which
govern the distension height $h_\mathrm{max}$, we
perform experiments and simulations on a single material while varying ring
radius $R$ or fluid depth $H$.
Similarly, to isolate material effects, we examine a single geometry
(fixed $R$ and $H$) while varying either the surfactant type or its
concentration. These experiments lie outside various of the model assumptions:
the lubrication approximation, insolubility, and monolayer concentrations.

\begin{figure}
\centerline{
\includegraphics[width=0.8\linewidth]{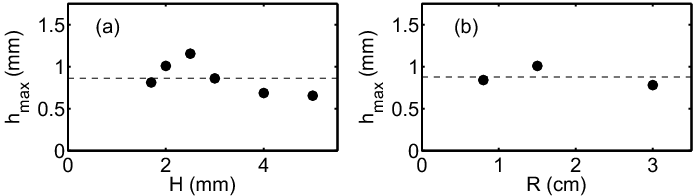}
}
\caption{{\bf Dependence of distension size $h_\mathrm{max}$ on geometric
parameters.}  LP experiments with a bulk concentration of PDMS.
($a$) For fixed $R=1.5$~cm and varying $H$ (second row of Table~\ref{t:Exper}), with $\langle h_\mathrm{max} \rangle =
0.86$~mm (dashed line). 
($b$) For fixed $H=2$~mm and varying $R$ (third row of Table~\ref{t:Exper}), with $\langle h_\mathrm{max} \rangle =
0.88$~mm (dashed line).
\label{f:hMax-Geometry}}
\end{figure}

\begin{figure}
\centerline{
\includegraphics[width=0.8\linewidth]{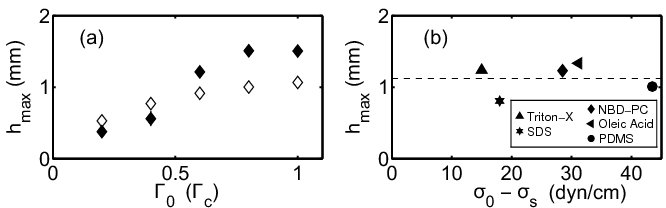}
}
\caption{{\bf Dependence of distension size $h_\mathrm{max}$ on material parameters.}
Simulations (open symbols) and both FI/LP and LP experiments (closed symbols).
($a$) Experiments and redimensionalized simulations with a variety of initial concentrations of NBD-PC and  fixed $H=0.7$~mm and $R=3.0$~cm (first row of Table~\ref{t:Exper}).
($b$) Experiments on five surfactant types, with fixed $R=1.5$~cm, $H=2.0$~mm,
and $V=540 \, \mu$L (fourth row of Table~\ref{t:Exper}). $\sigma_0 - \sigma_S$ 
quantifies the surface tension contrast for each surfactant.
Dashed line is $\langle h_\mathrm{max} \rangle = 1.12$~mm.
\label{f:hMax-dSigma}}
\end{figure}

First, we perform experiments using a bulk quantity of
PDMS as the surfactant, and probe the geometric effects using LP. We either 
vary the initial underlying glycerin thickness $H$ or the
retaining ring radius $R$ while leaving all other variables
constant. As shown in Fig.~\ref{f:hMax-Geometry}, we
observe a millimetric distension height in all cases
with no systematic dependence on either $H$ or $R$.

Second, we perform experiments and simulations examining the effects of different
surfactant materials and concentrations, using a combination of LP and FI/LP experiments. 
As shown in Fig.~\ref{f:hMax-dSigma}$a$, we vary the concentration of NBD-PC while leaving $R$ and
$H$ constant. We observe semi-quantitative agreement between the predicted and
observed values of $h_\mathrm{max}$ which increase as we increase $\Gamma_0$. In
Fig.~\ref{f:hMax-dSigma}$b$, we examine the full collection of bulk
surfactants, which includes soluble (Triton X-305, SDS), insoluble (PDMS, oleic
acid, NBD-PC), non-ionic (Triton X-305, PDMS), anionic (SDS, oleic acid), and zwitterionic (NBD-PC)
surfactants, using a fixed volume of surfactant in a fixed
geometry. In spite of this large parameter space, we find little dependence of
$h_\mathrm{max}$ on the surface tension difference. This may be due to the
existence of a large reservoir of bulk surfactant which is not present in
the monolayer experiments.

\section{Discussion and Conclusions}

In the axisymmetric self-healing of a surfactant layer on a thin fluid film,
gradients in the surfactant distribution generate Marangoni forces which drive
the fluid towards the center of the surfactant-free region.  The surface motion
of the fluid advects the surfactant, while the bulk motion of the incompressible
fluid pushes up an annular Marangoni ridge which coalesces into a central
distension.  This distension reaches a maximum height, after which it decays.

We approach this self-healing system in two ways.
First, we use a combination of fluorescence imaging and laser
profilometry to simultaneously probe the dynamics of both the fluid height and
surfactant concentration profiles. 
We find that while the commonly-used
model by \citet{Gaver1990}, together with the equation of state by \citet{Borgas-1988-MFT}, 
predicts reasonable fluid height
profile shapes, it does not accurately predict the spatial distribution
of surfactant. The agreement in the fluid height profile suggests the
validity of the lubrication approximation for this system.
In addition, the model correctly predicts the approximate shape of both
$r_M(t)$ and $r_S(t)$. Further, our measurements of $r_S$
give validity to Jensen's prediction \cite{Jensen1994} of $r_S
\propto (t_c - t)^{0.81}$.  At present, it remains unclear to what extent the
deviations in the surfactant concentration profile, the presence of a meniscus, and the
uncertainty in the glycerin viscosity each contribute to the
disagreement in the timescale.
Second, we probe the generality of the effect, beyond the model's
assumptions. Using laser profilometry, we  probe dependence of the distension
height on the initial fluid thickness, initial hole size, type of
surfactant, and surfactant concentration. In
all cases, the distension was millimetric in size, and only changes in the
concentration of surfactant were able to produce different $h_\mathrm{max}$.

In both these results and the droplet-spreading results of \citet{Swanson2013},
the experiments exhibited different behaviors at low surfactant concentrations:
the Marangoni ridge was suppressed, and the surfactant spreading drastically
slowed. In addition, simulations are not able to capture the sharp leading
edge of our inward-spreading front. These findings suggest that the
commonly-used linear or multilayer equations of state are insufficient to
capture low-$\Gamma$ dynamics \citep{Peterson2010}.  A promising future direction would be to
incorporate an empirically-determined equation of state.
Another interesting direction of investigation would be to probe the effect of the fluid meniscus created as the ring lifts. The observations presented in Fig.~\ref{f:Growth} indicate that surfactant molecules coalesce at the meniscus as the ring is slowly lifted, and leave behind an excess of surfactant that persists after pinch-off and propagates inward. For a linear equation of state, the incorporation of a meniscus and a small surfactant excess near the ring does not have a significant effect on the profiles or dynamics in  numerical simulations \citep{Swanson2013}. Again, an emperically-determined equation of state might provide additional insight into the real effects of the meniscus.

Based upon the lack of self-healing for dilute initial conditions (0.2
$\Gamma_c$), we conclude that there is either no Marangoni force present, or
that there are additional counterbalancing forces not accounted in the model.
For example, an equation of state with a vanishing derivative below a threshold
value of $\Gamma$ would produce no Marangoni force. For NBD-PC, our experiments
suggest that this threshold would be between $0.2$ and $0.4 \, \Gamma_c$.
Such an equation of state would additionally explain the sharpness of the
surfactant leading edge for
both droplet spreading \citep{Swanson2013} and self-healing.  In both cases,
when the concentration at the leading edge falls below this threshold, then the
Marangoni force at the leading edge vanishes and can no longer act to broaden
it. However, an equation of state with a vanishing derivative is likely not the 
complete story. As
discussed by \citet{Kaganer1999} for quasi-static systems, surfactant in sparse
concentrations undergoes a liquid-gas phase transition.  The sparse gas phase is
characterized by a two dimensional ideal gas law which stipulates that
$\sigma(\Gamma)$ should be linear. This provides a non-vanishing derivative for
the equation of state. It may therefore be necessary to also consider restoring
forces such as surface elasticity which are not present in
the \citet{Gaver1990} model.

The novel surfactant fluorescence visualization techniques developed in
\citep{Swanson2013} and used in this work have finally allowed us to make
quantitative comparisons to the well-accepted \citet{Gaver1990}  model   .  We see
clearly the success of this model in implementing the lubrication approximation,
and we have explained how this well-accepted model would benefit from incorporating a more
physically-motivated model of the surfactant monolayer.

\section{Acknowledgments}

This work was funded by the NSF grant DMS-FRG \#0968154 and the Research
Corporation Cottrell Scholar Award \#19788.  We thank Kali Allison for
conducting the initial experiments on this system, Jonathan Claridge and Jeffrey Wong for collaboration on the code, and Michael Shearer and
Ellen Swanson for insightful conversations.

\bibliographystyle{unsrtnat}
\bibliography{InwardRESUB}

\end{document}